**The Role of Molecular Shape on the Energetics of Chemisorption: From Simple to Complex Energy Landscapes**

David M. Huang[1] and Peter Harrowell[2,*]

[1]*School of Chemistry & Physics, The University of Adelaide, SA 5061, Australia*

[2]*School of Chemistry, University of Sydney, New South Wales, 2006, Australia*

**Abstract**

We enumerate all local minima of the energy landscape for model rigid adsorbates characterized by three or four equivalent binding sites (e.g. thiol groups) on a close-packed (111) surface of a face-centered-cubic crystal. We show that the number of energy minima increases linearly with molecular size, with a rate of increase that depends on the degree of registry between the molecule shape and the surface structure. The sparseness of energy minima and the large variations in the center-of-mass positions of these minima versus molecular size for molecules that are incommensurate with the surface suggests a strong coupling in these molecules between surface mobility and shape or size fluctuations resulting from molecular vibrations. We also find that the variation of the binding energy with respect to molecular size decreases more rapidly with molecular size for molecules with a higher degree of registry with the surface. This indicates that surface adsorption should be better able to distinguish molecules by size if the molecules are incommensurate with the surface.

---

[*] Corresponding author: peter.harrowell@sydney.edu.au

# 1. Introduction

The selective binding of molecules to well-defined binding sites is the physical basis of chemical communication and the biological recognition of molecules [1]. The relative binding energies of molecules to a crystal surface is a simpler example of this same selectivity and one of considerable interest in applications involving positioning and securing large functional molecules on a crystal surface. Common moieties in these fabricated systems are the (111) surface of gold and thiol groups [2]. The thiols can surrender the hydrogen, replacing the S-H bond with a S-Au bond. These sulfur linkages between a molecule and the gold have energies of 140 kJ/mol [3], a value easily large enough to qualify as chemisorption. Mobility on the surface can remain facile for molecules with multiple thiol groups, in spite of the strength of the molecule–surface interaction, due to the possibility of concerted attachment–detachment of different thiol groups [4].

A basic question concerning molecules bound to a surface via multiple thiol bonds is how does the binding energy of the molecule to the surface depend on the size, shape, and flexibility of the molecule? This question represents an important preliminary to understanding the dynamics of molecular adsorbates and the use of substrate binding to differentiate molecules. Both phenomena are currently the focus of considerable research activity. There are a number of experimental studies of the mobility of molecules on crystalline surfaces, predominantly the (111) surface of Cu or Pd [4,5]. Theoretical studies have explored the dynamics in a variety of models of the adsorbate-surface system [6]. A recent study [7] of the role of shape on the anisotropy of diffusion reports on the dynamics of the rigid molecule model employed in this paper.

Where chemical energetics is typically discussed in terms of a few well-characterized potential minima, systems with large numbers of local minima are generally discussed in terms of the statistics of the many local minima. This approach characterizes the studies of clusters [8], globular proteins [9], and glass-forming liquids [10]. This treatment is often referred to as the energy landscape approach. The explicit enumeration of the minima of the landscapes has been restricted to small atomic and molecular clusters [8]. The chemisorbed molecule represents a distinct class of systems in which the size of the molecule, relative to the lattice spacing of the surface, provides a physical control parameter that can continuously transform a simple energy landscape into a complex one.

In this paper we consider the dependence of the binding energy on a (111) surface of a molecule, characterized as a rigid polygon, each of whose $n$ vertices corresponds to a binding site, i.e. the $n$ thiol groups. The effect of bond flexibility (e.g. stretching, bending, or torsions) is not considered. Our aim here is not to reproduce the observed binding energy for a particular molecule but, rather, to use the simple model to efficiently explore generic trends in the dependence of the energy of adsorption on the size and shape of the molecular adsorbate.

## 2. Model and Methodology

The molecules in this model consist of $n$ sites connected by rigid bonds. These sites are the points of contact between the molecule and the surface and serve as the sole interaction with the surface. The total potential energy $V_n(\vec{r}^{\,n})$ of the molecule on the surface is

$$V_n(\vec{r}^{\,n}) = A_0 \sum_{i=1}^{n} \sum_{j=1}^{3} c_i \sin(\vec{k}_j \cdot \vec{r}_i) \,, \qquad (1)$$

where $\{\vec{k}_j\}$ are unit vectors that define the shape of the surface, $\vec{r}_i = (x_i, y_i)$ is the position of the $i$-th site on the surface, the $\{c_j\}$ are dimensionless constants, and $A_0$ is an arbitrary energy scale. For this work, we took

$$\vec{k}_1 = (1,0) \ ,$$
$$\vec{k}_2 = (\cos(\pi/3), \sin(\pi/3)) = (1/2, \sqrt{3}/2) \ ,$$
$$\vec{k}_3 = (\cos(-\pi/3), \sin(-\pi/3)) = (1/2, -\sqrt{3}/2) \,. \qquad (2)$$

With this choice of $\{\vec{k}_j\}$, $V_1(\vec{r})$ provides a simple representation of the interaction of a single site with the (111) surface of a face-centered-cubic (fcc) lattice (i.e. a two-dimensional triangular lattice) with a repeat length (2D lattice spacing) along any of the unit vectors of $a = 4\pi/\sqrt{3} \approx 7.255$. A contour plot of $V_1(\vec{r})$ is shown in Fig. 1.

This model makes several simplifying assumptions about surface binding. Firstly, it neglects motion of the molecule in the direction normal to the surface. This means that the model does not give the absolute binding energy of the molecule on the surface, but rather the relative binding energy of the interaction sites at different points on the surface. It therefore describes the position- and shape-dependent contributions to the binding energy. Secondly, the model does not account for different orientations of the molecule relative to the surface normal. Density functional calculations of $CH_3S$ on metal (111) surfaces have found that, although the absolute binding energy is sensitive to the angle between the C-S bond and the metal surface [11], the energy barrier to translational motion of a molecule bound to the surface varies only slightly with this angle [12]. Thirdly, although a more realistic molecular model would have included bond flexibility, only rigid molecules (i.e. molecules in which the positions of the interaction sites with respect to one another were kept constant) were considered in this work. The effects of molecular shape on surface binding and diffusion should be most pronounced in this case: adding bond flexibility would tend to smear out the influence of molecular geometry. In any case, bond flexibility is expected have a smaller impact on surface binding and diffusion than molecular shape for molecules without significant torsional degrees of freedom, such as the polycyclic aromatic hydrocarbons studied in Ref. [4]. This is because the variations in energy with molecular length $L$ or with displacement along the surface, when the energy and length scales in our model are matched to those for thiol chemisorption on metal surfaces [7], are significantly smaller than typical changes in energy with bond stretching and angle bending. On the other hand, energy barriers for torsions of atoms connected by single bonds are often only several $k_BT$, significantly smaller than variations in the energy landscape of surface binding sites. So molecules with torsional degrees of freedom, such as linear alkanes, can potentially maximize surface binding through bond rotations that change their conformation and thus their registry with the surface. For such molecules, however, the concept of molecular shape is ill-defined anyway. Fourthly, only the binding of a single molecule on the surface is considered. Thus, interactions between adsorbate molecules, which have been shown to have a strong effect on the binding energies of alkanethiols as a function of coverage on metal surfaces [3, 11], have been ignored. The results of this work are therefore relevant to the low-coverage regime. Although it is possible to extend the model to consider the impact of molecular shape on the energetics of chemisorption at high surface coverage, it is important to first understand how molecular shape affects binding without the complicating influence of interactions between adsorbate molecules.

Three different molecular shapes have been studied in this paper with interaction sites arranged as (a) an equilateral triangle with side length $L$, (b) an isosceles triangle with side lengths $L$, $L/\sqrt{2}$, and $L/\sqrt{2}$, and (c) a square with side length $L$. Sketches of these models are shown in Fig. 2.

For the rigid molecules, the positions of the interaction sites can be specified completely by the coordinates in the $(x,y)$-plane of a single point on the molecule and an orientational angle. We specify the position $\vec{r}_0 = (x_0, y_0)$ of the center-of-mass and the angle $\theta$ between the $y$-axis and the vector $\vec{r}_1 - \vec{r}_0$, where $\vec{r}_1$ is the position of site 1 (see Fig. 2). Defining $l_i = |\vec{r}_{i0}|$ and $\cos\theta_i = \dfrac{\vec{r}_{i0} \cdot \vec{r}_{10}}{|\vec{r}_{i0}||\vec{r}_{10}|}$, where $\vec{r}_{ij} = \vec{r}_i - \vec{r}_j$, the total potential energy can be written as

$$V_n\left(x_0, y_0, \theta\right) = A_0 \sum_{i=1}^{n} \sum_{j=1}^{3} c_i \sin\left(k_{jx}\left[x_0 + l_i \sin\left(\theta + \theta_i\right)\right] + k_{jy}\left[y_0 + l_i \cos\left(\theta + \theta_i\right)\right]\right). \qquad (3)$$

The diffusive motion of the molecules across the two-dimensional surface is very similar to what would be regarded in anthropomorphic terms as ``walking'' – a molecule moves by shifting some of its interaction sites while one or more of its other sites remain planted on the surface. Consequently, in what follows we refer to the molecules as ``walkers'' and their interaction sites as ``feet''. All distances and energies are measured in units of $a$ and $A_0$, respectively.

For the equilateral triangle (ET), isosceles triangle (IT), and square (SQ) walkers depicted in Fig. 2 in which all feet interact with the surfaces with equal strength ($c_i = 1$ in Eq. (3) for $i = 1, \ldots, n$), all local minima in the unit cell of the (111) surface were found. A combination of steepest-descent and conjugate-gradient (Polak–Ribiere) minimization methods [13] was employed, using analytical first derivatives of $V_n$ with respect to $x_0$, $y_0$, and $\theta$. The potential energy surface was searched exhaustively for local minima by running a series of minimizations with starting positions chosen randomly from a uniform distribution in the unit cell. An initial minimization was carried out by the steepest descent method with a relative step size in the coordinates of $10^{-2}$ and a relative convergence in the potential energy of $10^{-4}$. Further minimization by the conjugate gradient method was carried out to a relative convergence in the energy of $10^{-10}$. Minimizations were carried out until no new minima were found after 20,000 separate minimizations.

This method does not guarantee that all minima are found since some metastable minima can have basins of attraction that are very small, which may be missed if the steepest descent step size is too large. The initial bracketing of the minima and parabolic interpolation used to carry out line searches for minima along the conjugate directions in the conjugate gradient method may also miss minima [13]. However, due to the exhaustive nature of our searches (>20,000 minimizations from random starting positions in the unit cell for each molecule) and the identical results obtained with several different choices of the minimization parameters (steepest-descent step size and convergence tolerances for the energy in the initial steepest-descent minimization and subsequent conjugate-gradient minimization), we are confident that all minima were found with this method. Furthermore, in the case of the equilateral triangle molecule, the energies and molecular orientations of the minima exhibit particularly simple patterns, for which analytical results can be derived (see Results section and Supplemental Material), allowing the minimization algorithm to be verified. Any minima that may have been missed must also have such small basins of attraction to be of no physical relevance.

The minimization procedure was carried for side lengths $L$ (see Fig. 2) of the walkers between $a/10$ and $4a$ at intervals of $a/10$, where $a$ is the lattice spacing of the surface.

## 3. Results

### 3.1 The Cumulative Pattern of Energy Minima for the Equilateral Triangle

We shall first consider the case of a molecule with three feet arranged as in an equilateral triangle. A regular triangular molecule on a regular triangular lattice is a simple model. The energy landscape, however, quickly becomes quite complex as the separation $L$ between feet on the molecule increases relative to the lattice spacing. The distribution of energies of the local minima in a unit cell are shown in Fig. 3 for $L = 2a$, $4a$, and $8a$. All minima have the center-of-mass at one of three high symmetry sites: A $\left( \dfrac{a}{\sqrt{3}}, \dfrac{a}{2} \right)$, B $\left( \dfrac{\sqrt{3}a}{2}, a \right)$, or C $\left( \dfrac{2a}{\sqrt{3}}, \dfrac{3a}{2} \right)$, irrespective of the value of $L$. These sites are identified in Fig. 4. The developing complexity of the local minima is also evident in the apparently haphazard variation of the orientation of the molecule and the choice of sites (A, B or C) as the *global* minimum as $L$ is varied (as shown in Fig. 5). This complexity of ground states can give rise to similarly complex patterns in surface selectivity and surface mobility. Our aim is to understand the origin of these complex features of the local minima of the energy landscape as parameterized by the size and shape of the molecular adsorbate.

The energy of each of the three distinct minima is a roughly periodic function of $L$ with a period between $1.5a$ and $2a$ (see Fig. 6), although the oscillations in the energy only become strictly periodic in the limit of large $L$ (where they approach of period of $\sqrt{3}a$ ) [14]. As $L$ is increased beyond a specific length, $a$ for site A, $3a/2$ for B and $2a$ for C, the minima splits into two degenerate sets, differentiated by positive or negative angular deviations of equal magnitude about the original orientation [14]. The result is the degeneracy is doubled from 3 to 6. Fig. 6 also shows that new minima appear as $L$ increases; the energy of these minima is roughly periodic with $L$, but the oscillations are out of phase with those of the original minima. Collecting the global minima from the curves in Fig. 6, we have plotted in Fig. 7 the dependence of the lowest energy on $L$. This plot is characterized by the steady decay of the energy maxima with increasing $L$.

Turning to the lowest energy orientation of the molecule in each of the three distinct sites, we find a remarkable pattern of splitting and recurrence in each case, as shown in Fig. 8. For each of the three symmetry sites, the growth of the number of minima with increasing $L$ follows the same pattern: a pair of fan-like structures. Each fan has a base or root orientation – one at $\pi/6$ and the other at $\pi/2$ radians. For site A, the $\pi/6$ minimum appears first with respect to $L$. For site B, the $\pi/2$ minimum is found for the lowest values of $L$. Site C has no minima at very low values of $L$ but the $\pi/6$ minimum is the first to appear with increasing $L$. The fans have the same internal structure. The root minimum is stable for an interval in $L$ of $3a/2$ [14]. At the end of this interval, the root minimum becomes unstable as the feet of the molecule now find themselves on top of a saddle point in the surface potential. The root minimum then splits into a degenerate pair, as illustrated in Fig. 4. After a further increase in $L$ of $3a/2$, the root minimum returns, persists for an interval in $L$ of $3a/2$ and then splits again, and so on. The accumulation of the degenerate pairs of minima via this period doubling of the root minimum produces a step-wise linear increase the number of minima with increasing $L$. We note that the curvature of the orientation angle with respect to $L$ has opposite signs on the $\pi/6$ and the $\pi/2$ fans. The stationary points and cusps in the fans correspond respectively to global minimum energy configurations and configurations with energy $V = 0$.

As regular as the $L$ dependence of the number of minima associated with a single site is, the superposition of these three sites in determining the global minimum results in the complex sequence of orientations, shown in Fig. 5, that characterize the overall minimum with increasing $L$.

**3.2 Reduced Molecular Symmetry**

As a case study of the consequences of asymmetry in the molecule, we have determined the energy minima for an isosceles triangle arrangement of feet in which one side of the triangle is $\sqrt{2}$ times the length of the other two. The energies of the minima are plotted against $L$ in Fig. 9. Despite the shape difference, the isosceles triangle molecule landscape can be analyzed in the same way as the equilateral one. We find that there are three distinct types of minima corresponding to the A, B, and C sites from before, but unlike the equilateral triangle, the position of the center-of-mass is displaced from the surface symmetry site and this displacement changes with $L$.

We note that, relative to the equilateral molecule, (i) the number of minima at a given $L$ is reduced, (ii) the minimum energy is higher and the maximum energy is lower and (iii) the periodicity of the energies with respect to $L$ is gone with only a trace of the original undulations. Comparing the values of the energy at the global minimum vs $L$ for the isosceles molecule (see Fig. 10) with that of the equilateral molecule (Fig. 7), we find that the binding energy of the latter is consistently greater for all $L \geq 3a/2$. This means that our (111) surface is capable of energetically differentiating the two molecular shapes, irrespective of their relative sizes as long the values of $L$ for both molecules exceeds roughly the lattice spacing.

Figure 11 shows the orientation angle $\theta$ of the minima nearest to each of the three symmetry sites as $L$ is varied. The variation in $\theta$ shows familiar features of the regular 'fans' found for the equilateral triangle but now they occur as fragments only, for each of the different sites. In the case of the lower symmetry molecule, we now see the significant breaks in the minima already alluded to.

**3.3 Increasing the Number of Feet: the Square Molecule**

In going to the square molecule we increase the number of feet by one and so increase the possible binding energy. While we also increase the symmetry of the molecule, this is not a symmetry that is generally commensurate with that of the (111) surface. In Fig. 12, we plot the $L$ dependence of the energy minima of the square molecule about the three high-symmetry sites. We note that the breaks in the curves where no minimum was found for individual sites are even more marked than in the isosceles case. This means that the position of the square molecule on the surface when it adopts the lowest energy can depend sensitively on the value of the length $L$. This sensitivity raises the interesting possibility that for such systems there may be a very strong correlation between vibrations and mobility. In Fig. 13 we have overlaid the overall energy minima for the square molecule with that of the isosceles molecule. Here $L$ from Fig. 10 for the isosceles triangle molecule has been scaled by $1/\sqrt{2}$ so that the lengths of the two short sides are equal to the length of square. The similarity between the two curves is remarkable given the difference in the geometry and highlights a basic feature of shape recognition via chemical association. When differentiating molecules by their binding energies to a structured surface, it is the degree of registry between molecule and surface, rather than the actual molecular geometry, that is being measured.

## 4. Conclusions

In this paper we have examined how the energetics of molecular adsorption onto a crystalline surface depends on the shape and size of the molecule. The simple representation of the molecule–surface interaction is based on the use of thiol groups to anchor large molecules. We believe, however, that a number of our results are generic features of the symmetry and characteristic lengths of the molecule and the crystal surfaces and, therefore, may prove useful in thinking about molecular adsorption in general.

The binding energies of the molecules studied all exhibit a sequence of local minima with respect to the linear dimension $L$ of the molecule. The characteristic length of this sequence of minima was found to correspond to the distribution of distances between binding sites on the surface and to be roughly independent of the shape of the molecule itself. All molecules exhibit an increase in the number of minima as the size of the molecule increases. The rate of increase in the number of minima, however, depends on the degree of registry between the shape of the molecule and that of the binding surface. As the equilateral triangle molecule becomes larger, the number of minima increases linearly with $L$. This is graphically demonstrated in the nested fan structure of Fig. 8. For molecules with shapes that are generally not in registry with the surface structure, such as our isosceles triangle and square molecules, the growth in the number of minim with $L$ is considerably slower.

The equilateral triangle molecule is, of course, an atypical case. Its value to this study is to identify the full intricacies available to the energy landscape of the molecule–surface system. All molecules would be expected to deviate from this ideal, and, in so doing, restrict their space of minima, as we have seen in this study. Directly responsible for this 'fusing' of minima is the coupling between the size and center-of-mass position in the energy minima of non-equilateral molecules. This coupling suggests that center-of-mass mobility can involve a strong coupling to molecular vibrations that see fluctuations in the molecular shape and size, especially for average values of $L$ close to points where minima overlap.

How effective can a simple geometry such as that of a closed-packed crystal surface be in differentiating between molecular size and shape on the basis of their binding energies? As far as shape discrimination goes, we see the most striking differences in binding energy when we compare a molecule that is commensurate with the surface (such as the equilateral triangle, see Fig. 7) and one of the shapes that is not in close registry with the surface (i.e. Fig. 10 or Fig. 13). The difference in binding energies for different shapes is a sensitive function of $L$. Our surface's capacity to distinguish between two shapes with little registry with the surface structure is considerably poorer. In terms of differentiating between molecules of the same shape but different size, we find that the variation in the binding energy with respect to $L$ decays quickly with increasing molecular size. This decay is somewhat slower for the lower symmetry molecules. The question as to whether there are other simply defined surface structures that can improve the differentiation of adsorbate shape (over a limited size range) remains an interesting open question.

**Acknowledgments**

We would like to acknowledge valuable discussion with David Barret, whose unpublished study of this problem (Chemistry Honours 2004, University of Sydney) provided considerable guidance for this present study.

**Figures**

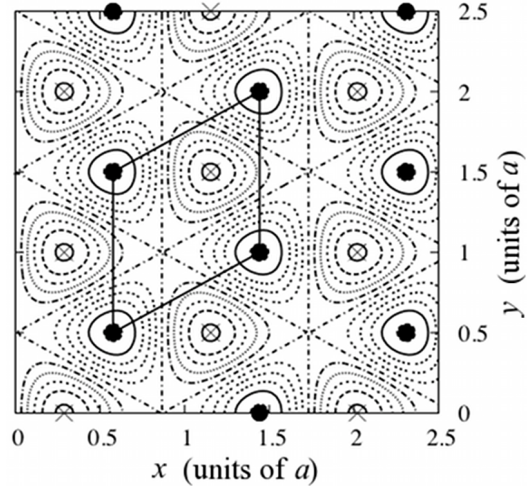

FIG. 1. Potential energy $V_1(\vec{r}) = V_1(x, y)$ of a single site with interaction strength $c_1 = 1$ on the (111) surface. The minimum energy (circles) is $-3\sqrt{3}A_0/2 \approx -2.598A_0$ and occurs when $(x, y) = \left(\dfrac{(3m - 1/2)a}{\sqrt{3}}, na\right)$ and $\left(\dfrac{(3m + 1)a}{\sqrt{3}}, (n + 1/2)a\right)$, where $m$ and $n$ are integers. The maximum energy (crosses) is $3\sqrt{3}A_0/2 \approx 2.598A_0$ and occurs when $(x, y) = \left(\dfrac{(3m + 1/2)a}{\sqrt{3}}, na\right)$ and $\left(\dfrac{(3m - 1)a}{\sqrt{3}}, (n + 1/2)a\right)$. The potential is zero along the lines $x = na\sqrt{3}/2$ and $y = (n + 1/2)a \pm mx/\sqrt{3}$. The solid line shown on the surface surrounds the unit cell of the lattice bounded by minima at $(a/\sqrt{3}, a/2)$, $(a/\sqrt{3}, 3a/2)$, $\left(5a/\left(2\sqrt{3}\right), a\right)$ and $\left(5a/\left(2\sqrt{3}\right), 2a\right)$ and containing a maximum at $(2a/\sqrt{3}, 3a/2)$.

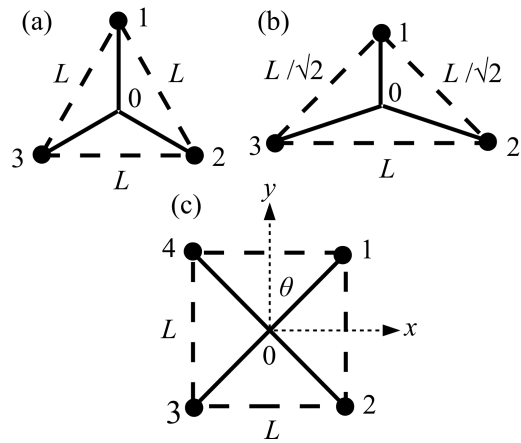

FIG. 2. Pictures of the three molecular shapes – (a) equilateral triangle, (b) isosceles triangle, and (c) square – with the length $L$ and center-of-mass "0" indicated on each. As indicated in (c), the molecular orientation is defined by the angle $\theta$ between the $y$-axis and the vector $\vec{r}_1 - \vec{r}_0$ joining the center-of-mass to site 1.

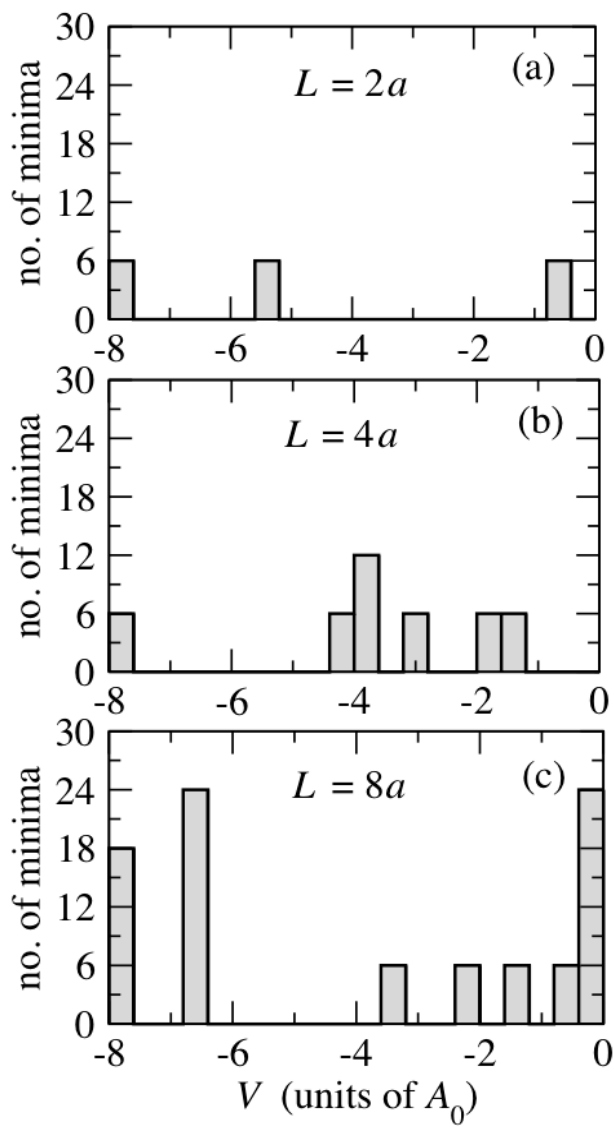

FIG. 3. Bar charts of the number of global minima in a single unit cell as a function of energy for the equilateral triangle molecule with $L = 2a$, $4a$, and $8a$.

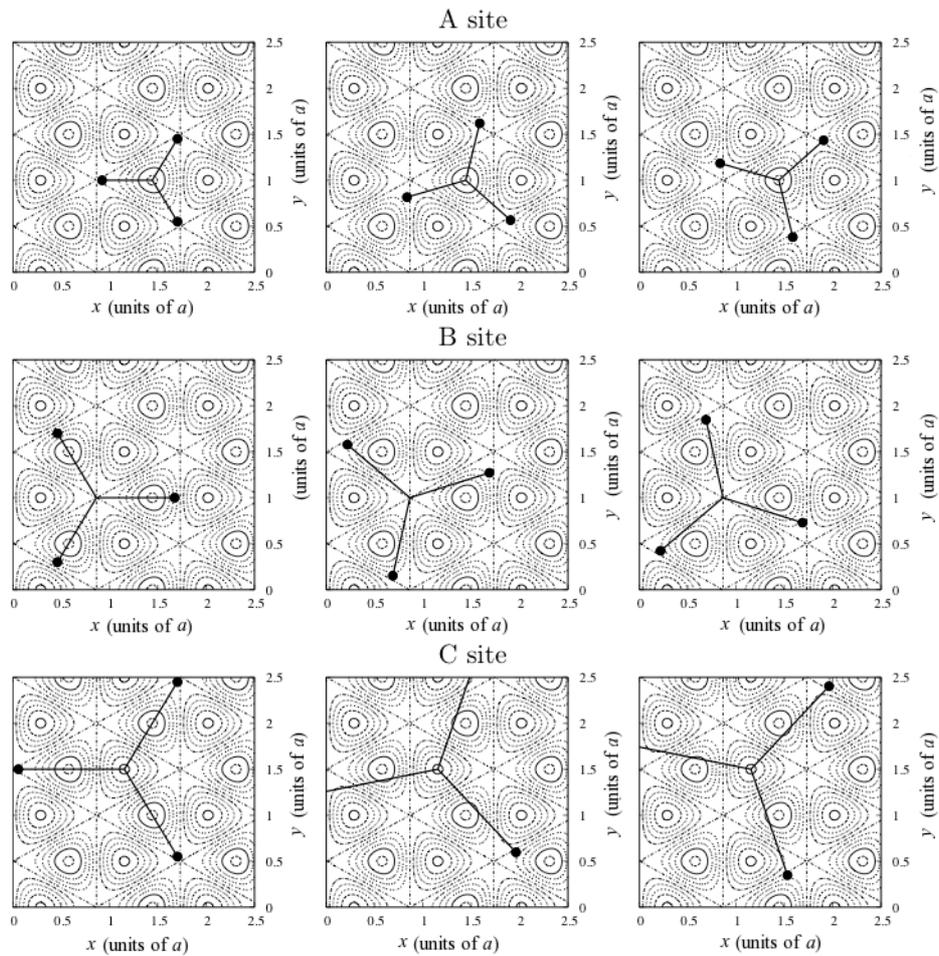

FIG. 4. The equilateral triangle molecule at the value of $L$ at which the minima split, showing both the original orientation (left panel), lying on a symmetry point, and the two new degenerate orientations (center and right panel), for each of the three sites, A (top: splitting at $L = a$), B (middle: splitting at $L = 3a/2$) and C (bottom: splitting at $L = 2a$).

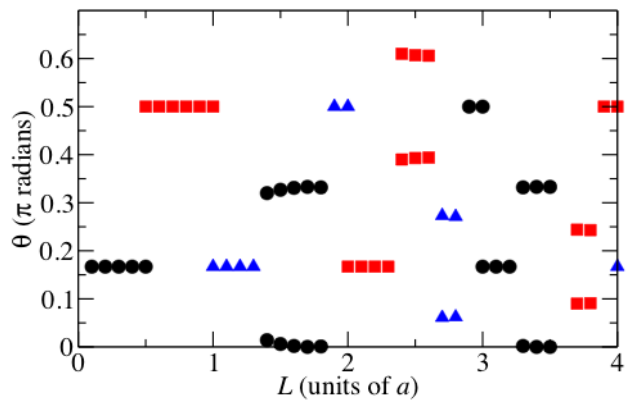

FIG. 5. (Color online) Angle $\theta$ between one arm of the equilateral triangle molecule and the $y$ axis at the global minimum or minima for $0 \leq L \leq 4a$. (Only 1/3 of the minima are depicted in this and subsequent figures; the rest can be obtained by adding $2\pi/3$ and $4\pi/3$ respectively to the values of $\theta$ depicted.)

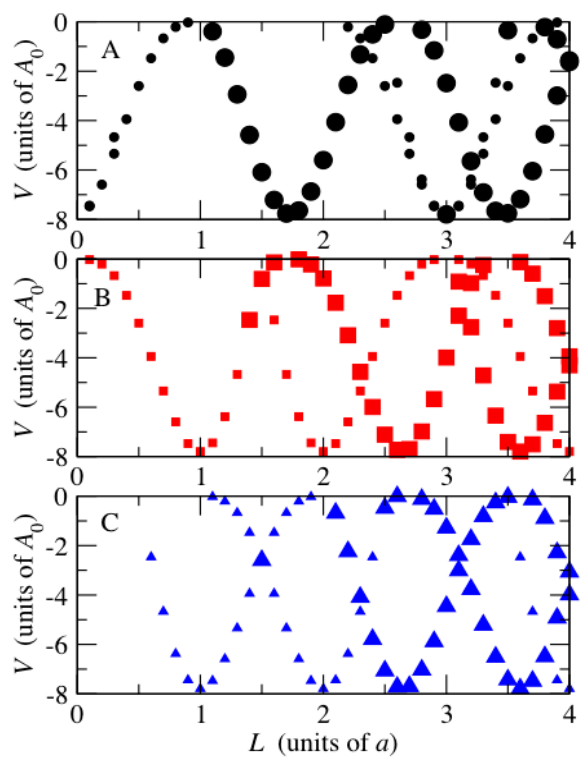

FIG. 6. (Color online) Energy vs length $L$ for minima at each of the three sites for the equilateral triangle molecule. The small and large symbols represent three-fold and 6-fold degenerate minima respectively.

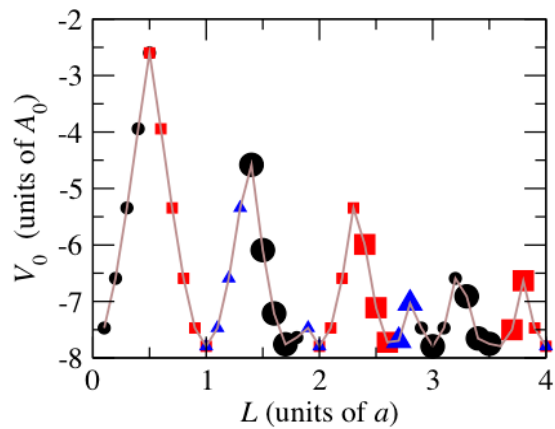

FIG. 7. (Color online) Combination of the data in Fig. 6 showing only the overall

ground states.

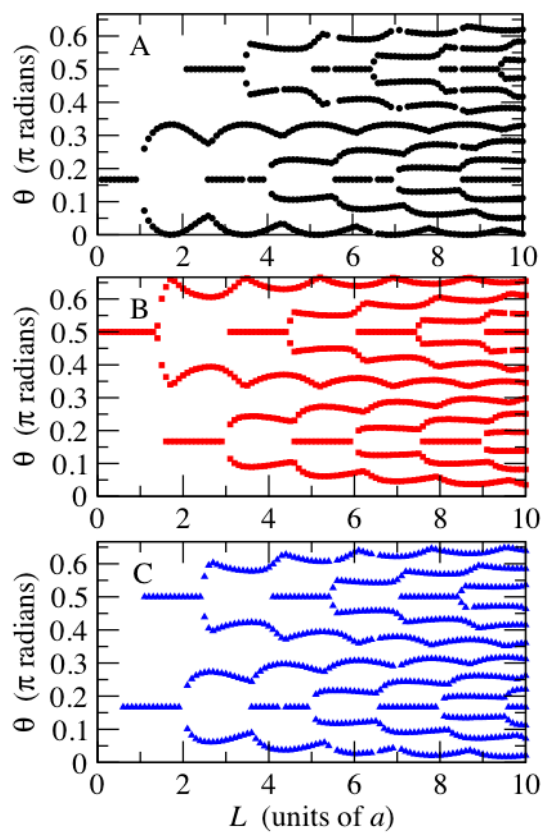

FIG. 8. (Color online) Orientational angle $\theta$ of the local minima vs length $L$ for $0 \leq L$

$\leq 10a$ for the three sites of the equilateral triangle molecule.

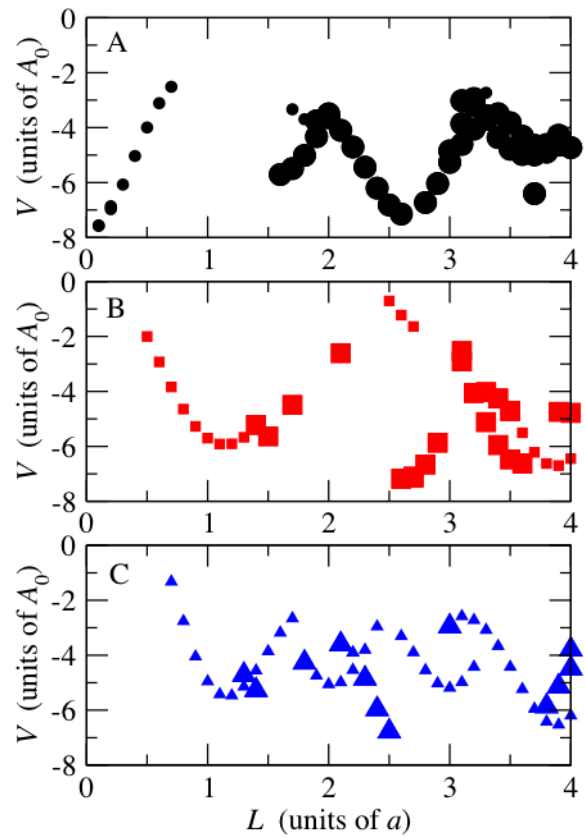

FIG. 9. (Color online) Energy vs length $L$ for minima closest to each of the three high-symmetry sites for the isosceles triangle molecule.

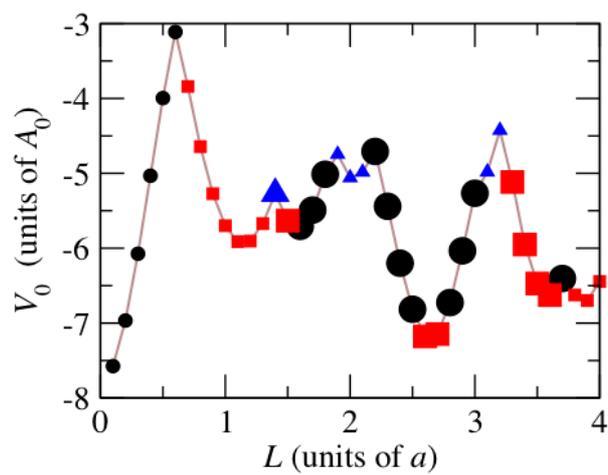

FIG. 10. (Color online) Combination of the data in Fig. 9 showing the overall ground states for the isosceles triangle molecule.

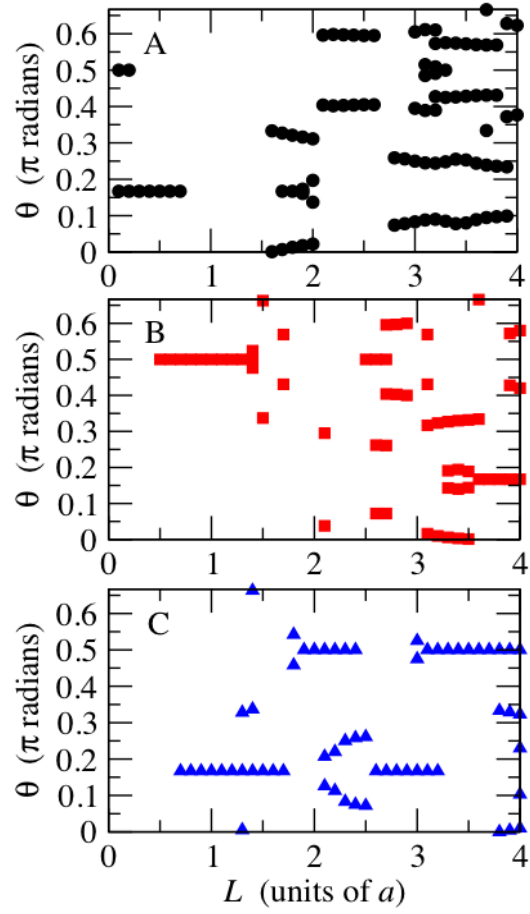

FIG. 11. (Color online) Orientational angle $\theta$ of the local minima vs length $L$ for $0 \leq L \leq 4a$ for the three sites of the isosceles triangle molecule.

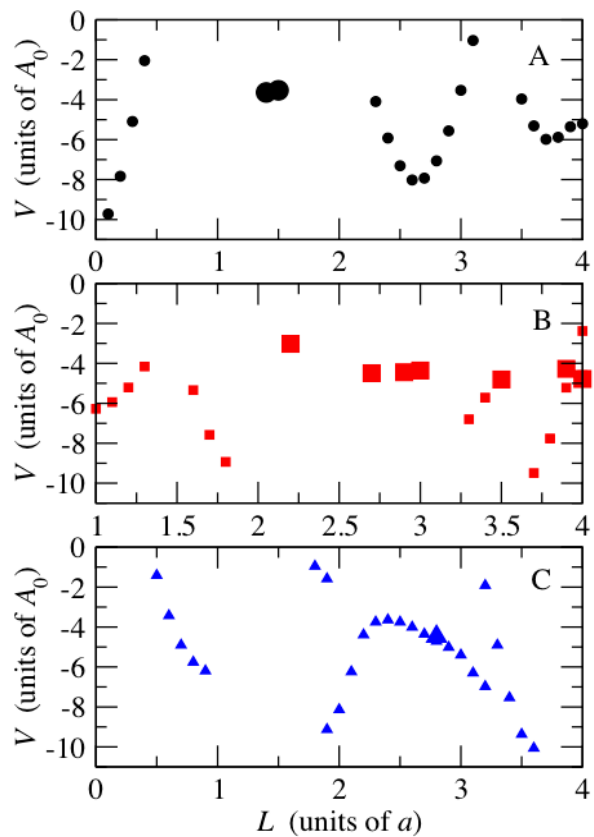

FIG. 12. (Color online) Energy vs length $L$ for minima closest to each of the three sites for the square molecule.

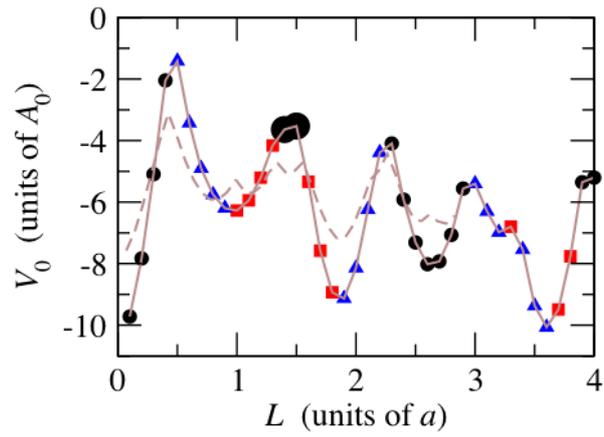

FIG. 13. (Color online) The overall minimum energy for the adsorbed square molecule (solid line) corresponding to the lower envelope of the data from Fig. 12. For comparison, the analogous lowest energy for the isosceles triangle (i.e. the data from Fig. 10) is plotted (dashed line) with $L$ scaled by $1/\sqrt{2}$ .